\documentclass[12pt,twoside]{article}   

\usepackage{fancyhdr}		

\usepackage[sorting=none]{biblatex}
\usepackage[section]{placeins}   %

\usepackage{graphicx}

\makeatletter \renewcommand\@biblabel[1]{$^{#1}$} \makeatother
\newcommand{\cen}[1]{\begin{center} #1 \end{center}}

\usepackage{hyperref}
\hypersetup{ colorlinks,
    citecolor=blue,
    filecolor=blue,
    linkcolor=blue,
    urlcolor=blue
}

\usepackage{xcolor}

\definecolor{gray}{rgb}{0.6,0.6,0.6}
\definecolor{red}{rgb}{0.85,0,0}
\definecolor{green}{rgb}{0,0.85,0}
\definecolor{blue}{rgb}{0,0,0.85}
\definecolor{beige}{rgb}{0.92,0.87,0.78}
\usepackage[all]{hypcap}    

\begin{document}

\cen{\sf {\Large {\bfseries Quantitative real-time measurements of dose and dose rate in UHDR proton pencil beams via scintillation imaging system} \\  
\vspace*{10mm}
Megan Clark\textsuperscript{1}, Joseph Harms\textsuperscript{2}, Roman Vasyltsiv\textsuperscript{1}, Austin Sloop\textsuperscript{1}, Jakub Kozelka\textsuperscript{3}, Bill Simon\textsuperscript{3}, Rongxiao Zhang\textsuperscript{1}, David Gladstone\textsuperscript{1}, Petr Bruza\textsuperscript{1}}\\

\textsuperscript{1}Thayer School of Engineering, Dartmouth College, Hanover, NH \\
\textsuperscript{2}University of Alabama at Birmingham, Department of Radiation Oncology, Birmingham, AL\\
\textsuperscript{3}Sun Nuclear Inc., Melbourne FL 32940, US \\
\vspace{5mm}
Version typeset \today
}

\pagenumbering{roman}
\setcounter{page}{1}
\pagestyle{plain}
\cen{Author to whom correspondence should be addressed is Megan Clark.\\
email: megan.a.clark.th@dartmouth.edu}

\newpage
\begin{abstract}
\noindent {\bf Background:} Ultra-high dose rate radiotherapy (UHDR-RT) has demonstrated normal tissue sparing capabilities, termed the FLASH effect; however, available dosimetry tools make it challenging to characterize the UHDR beams with sufficiently high concurrent spatial and temporal resolution. Novel dosimeters are needed for safe clinical implementation and improved understanding of the effect of  UHDR-RT.\\ 
{\bf Purpose:} Ultra-fast scintillation imaging has been shown to provide a unique tool for spatio-temporal dosimetry of conventional cyclotron and synchrocyclotron pencil beam scanning (PBS) deliveries, indicating the potential use for characterization of ultra-UHDR PBS proton beams. The goal of this work is to introduce this novel concept and demonstrate its capabilities in recording complex dose rate maps at FLASH-capable proton beam currents, as compared to log-based dose rate calculation, internally developed UHDR beam simulation, and a fast point detector (EDGE diode). \\
{\bf Methods:} The light response of a scintillator sheet located at isocenter and irradiated by pencil beam scanning proton fields (40-210 nA, 250 MeV) was imaged by an ultra-fast iCMOS camera at 4.5-12 kHz sampling frequency. Camera sensor and image intensifier gain were optimized to maximize the dynamic range; the camera acquisition rate was also varied to evaluate the optimal sampling frequency. Large field delivery enabled flat field acquisition for evaluation of system response homogeneity. Image intensity was calibrated to dose with film and the recorded spatio-temporal data was compared to a PPC05 ion chamber, log-based reconstruction, and EDGE diode. Dose and dose rate linearity studies were performed to evaluate agreement under various beam conditions. Calculation of full-field mean and PBS dose rate maps were calculated to highlight the importance of high resolution, full-field information in UHDR studies.\\
{\bf Results:}Camera response was linear with dose (R\textsuperscript{2} = 0.997) and current (R2\textsuperscript{2} = 0.98) in the range from 2-22 Gy and 40-210 nA, respectively, when compared to ion chamber readings. The deviation of total irradiation time calculated with the imaging system from the log file recordings decreased from 0.07\% to 0.03\% when imaging at 12 kfps versus 4.5 kfps. Planned and delivered spot positions agreed within 0.2 ± 0.1 mm and total irradiation time agreed within 0.2 ± 0.2 ms when compared with the log files, indicating the high concurrent spatial and temporal resolution. For all deliveries, the PBS dose rate measured at the diode location agreed between the imaging and the diode within 3 ± 2\% and with the simulation within 5 ± 3\% \\
{\bf Conclusions:} Full-field mapping of dose and dose rate is imperative for complete understanding of UHDR PBS proton dose delivery. The high linearity and various spatiotemporal metric reporting capabilities confirm the continued use of this camera system for UHDR beam characterization, especially for spatially resolved dose rate information. \\

\end{abstract}

\newpage
\setlength{\baselineskip}{0.7cm}      
\pagenumbering{arabic}
\setcounter{page}{1}
\pagestyle{fancy}
\section{Introduction}
\indent Proton therapy delivered at ultra-high dose rates has indicated a potential reduction in normal tissue toxicity\cite{diffenderfer_current_2022,favaudon_ultrahigh_2014,konradsson_correction_2020,sorensen_pencil_2022,velalopoulou_flash_2021,vozenin_advantage_2019}, termed the FLASH effect, leading to the first in-human clinical trial: The FAST-01 Nonrandomized Trial \cite{mascia_proton_2023}. Some studies suggest a dose-rate threshold of 40-100 Gy/s is needed to elicit this effect, thus requiring that treatment planning considers dose rate in addition to traditional dose volume histogram-based planning \cite{rothwell_treatment_2022, wei_novel_2021}. Despite the rapid push towards UHDR clinical implementation, currently available dosimetry tools make it challenging to fully characterize and monitor the dose and dose rates being delivered \cite{taylor_roadmap_2022}.\\
\indent Ionization chambers used in proton FLASH studies, namely the PPC05 parallel-plate ionization chamber, exhibit dose rate dependencies, thus requiring extensive calibration of the recombination factor \cite{konradsson_correction_2020,kranzer_ion_2021,lee_ultrahigh_2022,mcmanus_challenge_2020}. Additionally, ionization chambers only provide information on a small sensitive volume (for PPC05 0.05 cm\textsuperscript{2}), making dose-rate validation of the entire scanned field impossible. A 2D strip ionization chamber has reported strong spatial-temporal mapping of UHDR proton PBS deliveries at a sampling rate of 20 kHz, but requires implementation of 2D dose reconstructions and a fixed geometry setup \cite{yang_2d_2022}. EDGE detector diodes and scintillator fibers have demonstrated dose and dose rate (mean and instantaneous) independence at sub millisecond resolution but lack concurrent spatial information \cite{kanouta_time_2022,levin_scintillator_2023,rahman_characterization_2023}. Other novel detection methods include PET imaging and silicon carbon detectors, which suffer from similar challenges \cite{abouzahr_first_2023,romano_first_2023}. In summary, there are no existing devices that allow for full-field spatial resolution at the temporal resolution required for real-time monitoring of UHDR beams. \\
\indent Luminescent detectors have been shown to provide adequate spatial and temporal resolution for the demands of UHDR-RT \cite{ashraf_dosimetry_2020,beaulieu_review_2016,el_naqa_image_2022,verhaegen_future_2020,yogo_luminescence_2023}. Specifically, scintillation-imaging has been used to acquire accurate dose and dose rate measurements of clinical cyclotron and synchrocyclotrons, indicating potential use for high-speed mapping of the scanning beam \cite{clark_ultra-fast_2023,rahman_utilizing_2022}. It also enables remote monitoring of the beam during delivery and accommodation of different imaging surfaces and geometric setups. Advancements to the camera system previously described by Clark et al, allow for over 12 kHz frame rate acquisitions, thus extending the application of this imaging system to ultra-high dose rate capable PBS systems \cite{clark_ultra-fast_2023}. \\
\indent The intensified camera used in this study is the first-of-its-kind to image an UHDR cyclotron beam at up to 12 kHz frame rate and with concurrent submillimeter (0.22×0.22 mm\textsuperscript{2}) spatial resolution. In this study, the camera sensor and image intensifier gain were optimized to maximize the dynamic range, and the camera acquisition rate was varied to evaluate the optimal sampling frequency. Image intensity was calibrated to dose with film and camera angular corrections, flat and dark field, and background subtraction image corrections were performed in MATLAB. The recorded spatio-temporal data was compared to a PPC05 ion chamber, internally developed simulation, and EDGE diode for validation of the reconstruction model and evaluation of the imaging system performance. Dose and dose rate linearity studies were performed to evaluate agreement under various beam conditions. Calculation of full-field mean, instantaneous, and PBS dose rate maps were calculated to highlight the importance of high resolution, full-field information in UHDR studies and demonstrated the importance of various critical reporting parameters for UHDR PBS deliveries.

\section{Methods}
\subsection{Experimental Setup}
\indent An intensified CMOS camera (BeamSite UTLRA, DoseOptics) was set up on a tripod to image the light response of scintillator sheet during irradiation of pencil beam scanning proton fields (30-99 nA, 250 MeV) in the treatment vault of a Varian ProBeam cyclotron system at the University of Alabama Birmingham Proton Center in Birmingham, Alabama, Figure \ref{figure1}. The acquisition frame rated ranged from 4.5-12 kHz, and all frames were streamed and saved to a fast computer solid state drive (SSD). A Gen2 blue-sensitive intensifier with P46 fast-decay phosphor was specifically chosen to avoid temporal blur due to luminescence decay and afterglow in more standard phosphors. Camera analog gain of 1V and image intensifier gain of approximately 1.8V were selected to maximize the dynamic range. The camera acquisition rate was varied to evaluate the impact on dose rate calculations and to optimize the sampling frequency used in later studies. Large field delivery enabled flat field acquisition for evaluation of spatial homogeneity.\\
\indent A beam target consisting of an 8x11 in, 0.2 mm thick scintillator sheet (Rapidex by Scintacor, Cambridge UK) placed at isocenter and laid over stacks of solid water. Image intensity was calibrated to dose with film (Gafchromic EBTXD, Ashland), which was placed and aligned directly under the scintillator sheet. An EDGE diode detector and PPC05 ion chamber were placed 5 cm and 3 cm below the film for temporal and cumulative dose comparisons, respectively. When operating in UHDR mode, the VarianProbeam log file system only reports cumulative dose, total irradiation time, dose rate, and percent of the field receiving a certain dose rate. There is limited publicly available information regarding the parameters required for accurate calculation of dose rate, as is done by the log file system. Therefore, an internally developed simulation (Section \ref{simsection}) was developed in MATLAB for a more comprehensive comparison to the imaging data. One of the secondary aims of this work was to validate this beam simulation for improved treatment planning of UHDR deliveries at UAB. 

\subsection{Treatment Plans}
\indent The baseline treatment plan for this study was designed to deliver 15 Gy to the treatment isocenter with a spot spacing of 5 mm and single spot dwell times around 3-4 ms, based on recommendations from the vendor. A diamond pattern was used to maximize PBS dose rate, which will be defined in Section \ref{subdoserate} below. The fields were 5 cm across the central axis in both x and y to ensure lateral charged particle equilibrium at the center of the field for accurate verification with an independent ion chamber. To characterize linearity of the camera and scintillation system response with varying dose, the MU/spot were either decreased or increased from the nominal plan to deliver fields from 2 Gy up to 22 Gy. Dose rate linearity was studied by delivering the 15 Gy baseline plan with various scanning nozzle currents ranging from 30 nA to 99 nA. The system responded linearly with dose and current (R\textsuperscript{2} = 0.997, R\textsuperscript{2} = 0.98), and so these fields were used for validation of the simulation, as shown in Figure \ref{figure1}. UHDR PBS fields were delivered with a Varian Probeam system converted to ultra-high dose rate mode after clinical hours. 
\subsection{Image Processing}
\indent All image processing was performed in MATLAB (Version 2022b). Image stacks were offset-corrected with background subtraction, divided by the flat-field image for spatial and camera nonlinearity correction, spatially transformed and pixel-size calibrated to account for the camera angle, and the image intensity was calibrated to absolute dose with film \cite{clark_ultra-fast_2023}. Each individual image of the scanning spot was used for evaluation of beam position, dose, and dose rate. To identify the center of the scanning spot, each frame was fit to a gaussian and the center of the gaussian was localized for comparison to the treatment plan. \\
\indent Following the image correction methods previously defined by Clark et al in 2022 for imaging at 1 kHz of a conventional synchrocyclotron, the camera was focused to a spatial calibration grid located at isocenter under traditional room lighting and low intensifier gain \cite{clark_ultra-fast_2023}. Imaging of this calibration grid provided a geometric transform (to account for the oblique angle of the camera relative to the scintillating sheet/incident beam), pixel size calibration, and spatial resolution measurements to be used in image analysis. A large, 10×20 cm\textsuperscript{2} (flat) field was delivered at 250 MeV and imaged at 4.5 kHz concurrently with EBTXD film measurement located under the scintillator sheet. A flat-field correction map was calculated by dividing the geometry-corrected cumulative camera image by the 2D dose distribution measured by film. All subsequent image stacks were divided by this correction map to minimize spatial response inhomogeneity of the scintillator and camera image forming components, including lens and intensifier. 
\subsection{Frame Rate Optimization}
\begin{figure}
\begin{center}
\includegraphics[width=16.5cm]{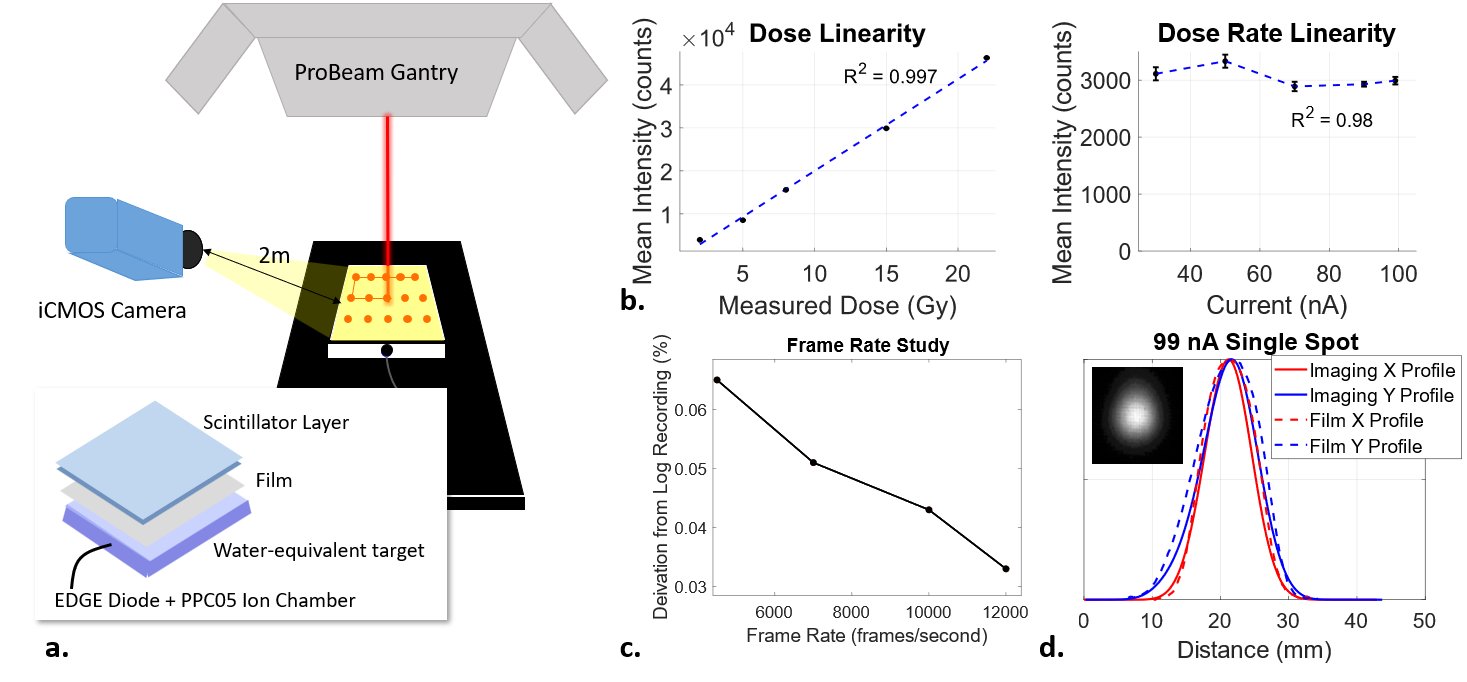}
\label{figure1}\caption{ a.) Camera facing scintillator sheet, diode, and PPC05 at isocenter. b.) Results of dose and dose rate linearity studies from 2-22Gy and 30-220nA, respectively. c.) Impact of frame rate on imaged dose rate deviation from log file. d.) Agreement between imaged and film data of single spot profile at 99nA.}

\end{center}
\end{figure}

\indent A single spot plan was delivered and imaged at 10, 50, 70, and 99 nA with film, acting as the reference, to offer insight into camera spatial accuracy. The imaging system provided 0.22×0.22 mm\textsuperscript{2} spatial resolution, such that the 99 nA single spot full width half maximum agreed with the film within 0.5 mm in both the x and y directions. The intensifier and sensor electronic gains were set and kept constant for this entire study to accommodate high current deliveries, as ultra-high dose rates were the beams of interest; this leads to the higher dose rate linearity error and spot disagreement at low currents. To find the optimal combination of electronic gains, we repeatedly imaged a 99 nA beam delivery while varying the intensifier gain (IG) from 3V down to 0.8 V and sensor gain (“analog” gain, AG) from 1x to 4x. At low IG/high AG the amplified sensor noise limited the low photon response, while at extremely high IG we observed temporary (millisecond time scale) charge depletion leading to locally decreased sensitivity of the camera. An optimal combination of IG = 1.8V and AG = 1V was set and kept constant for all measurements here.\\
\indent This updated camera system can image at over 12,000 frames per second. A frame rate study was performed by imaging the same beam at 4.5k, 7k, 10k, and 12k to investigate improvements in temporal accuracy with higher frame rates. For all frame rates, deviation from the log files was under 0.32 ms, with deviation from log file total irradiation time decreasing with increasing frame rate (Figure \ref{figure1}). This highlighted the potential capabilities of the camera system for collecting accurate dose rate information, but a 4.5kHz frame rate was chosen for this study to capture sufficiently accurate per-spot information while minimizing data size for processing and storage purposes. 

\subsection{Simulation and Log-file Comparison} \label{simsection}
\indent To acquire a better understanding of the dynamics of UHDR PBS deliveries, an independent simulation was developed in MATLAB to mimic machine delivery and provide further validation of the scintillator imaging system. This simulation takes as input either a measured single spot map or functional spot model (i.e. a 2D Gaussian distribution), planned spot map including MU/spot, and scanning nozzle current. The user can also input parameters dose, dose rate, and thresholds for calculation of PBS dose rate. The output of the simulation are dose and dose rate maps for the field in question. From these maps, parameters of interest such as mean dose rate or PBS dose rate for a given pixel can be extracted. For initial validation, the total delivery time for each field was used to determine scanning speed in the X and Y directions. For comparison to the simulation, gamma analysis was performed on the full field images at 3\%/2mm, per proton characterization standards.  
\subsection{Dose Rate Definitions} \label{subdoserate}
\indent Unlike point detectors, this imaging system enables full-field implementation of different dose rate definitions. Pencil beam scanning dose rate, $D_{PBS}$, defined by Folkerts et al. in 2021, was used for comparison of the diode, log-based reconstruction, and log file recordings, as per convention in UHDR proton beam studies and the FAST-01 trial \cite{mascia_proton_2023, folkerts_framework_2020, }. This dose rate definition only considers the amount of time that dose, within a certain dose threshold $d_{\dagger}$ of the min/max dose $D(\vec{x})$, is delivered to a certain point in the field, $\vec{x}$: \\
\begin{equation} \label{eq:1}
\dot{D}_{PBS}(\vec{x})=\frac{D(\vec{x})-2d_{\dagger}}{T(\vec{x})}.
\end{equation}
\indent In addition to PBS dose rate maps, full-field maps of mean and maximum instantaneous dose rate were calculated to highlight the variability of dose rate with location in the field and definition of dose rate used. The mean dose rate to a point $\vec{x}$ is defined as the total dose to that point, $D(\vec{x})$, divided by the total delivery time, ${T_{tot}}$:
\begin{equation} \label{eq:2}
\dot{D}_{mean}(\vec{x})=\frac{D(\vec{x})}{T_{tot}} .
\end{equation}		
The maximum instantaneous dose rate is defined as the maximum dose to a point in the field over the time between sampling frames: $\dot{D}_{inst}=\frac{D_{max}(\vec{x})}{dt}$.\\
\indent As is further discussed in the results section, reporting of dose thresholds used for PBS dose rate definitions is imperative for meaningful comparison of dose rate. For this work, a threshold dose, $d_{\dagger}$ of 0.5 Gy was chosen and held constant for image analysis of all deliveries. While Varian reports a $d_{\dagger}$ equal to 0.01 Gy  in the calculation of PBS dose rate, the limited dynamic range of the camera, which leads to uncertainty in the low-dose region of each spot, required the higher chosen threshold. An additional 10\% maximum dose threshold was applied when calculating the percent of the field greater than 40Gy/s, such that only regions of the field receiving greater than 10\% of the maximum dose were considered. Investigation into the impact of $d_{\dagger}$ on the calculated dose rate is reported and compared to that simulated in the original publication by Folkerts. Additional study into the impact on percent of the field greater than 40\% is provided in the results here. While Folkerts et al. simulate the dependency on dose rate on $d_{\dagger}$, we believe this is the first measured dose rate area histogram demonstrating this relationship. 
\\

\section{Results}
\subsection{Spot Tracking and Beam Dynamics Characterization}
\begin{figure}
\begin{center}
\includegraphics[width=15cm]{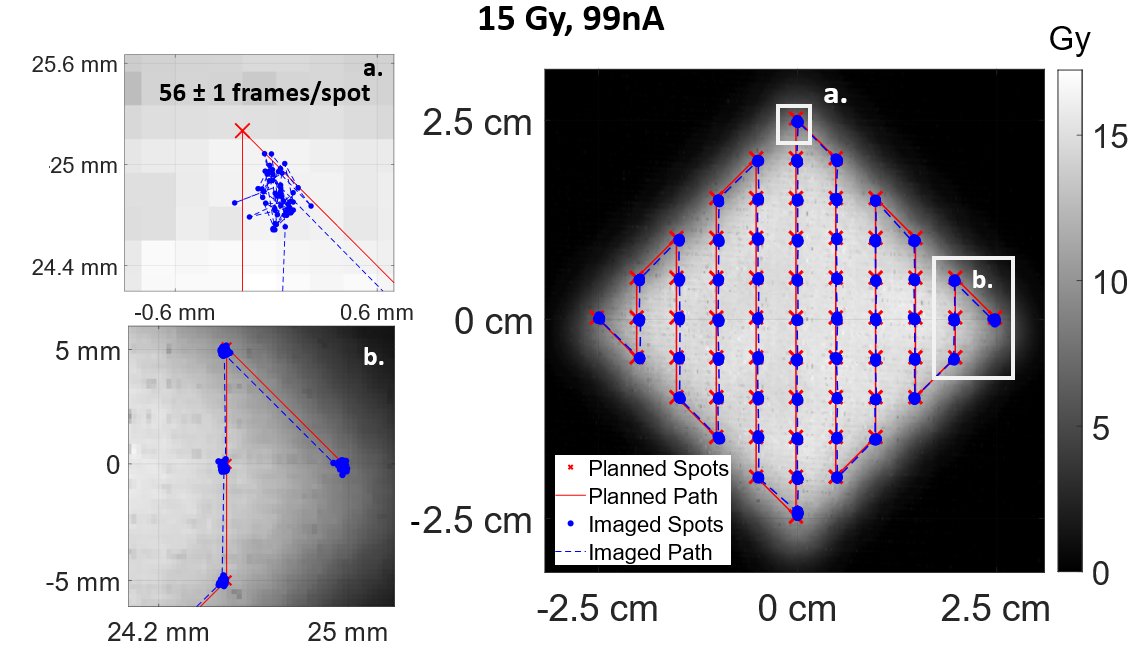}
\caption{Illustration of scanning spot localization agreement to plan and ability to capture multiple frames per spot.}
\label{fig_example2} 
\end{center}
\end{figure}

An acquisition rate of 4,500 frames/second enabled accurate estimates of dynamic scanning spot parameters including spot positions, spot dwell time, spot scanning speed, and total irradiation time. Imaged spot positions agreed within 0.2 ± 0.1 mm to the planned spot positions (Figure \ref{fig_example2}). Spot dwell time increased from 3.61 ms for the 2 Gy delivery to 56.2 ms for the 15 Gy delivery, with an average standard deviation in dwell time of only 0.2 ms. A 5x5 cm\textsuperscript{2} field was delivered with increased, 8 mm, spot spacing to enable improved estimating of spot scanning speeds. This provided measurements of spot scanning speeds of 12 ± 1 m/s in the x direction and 24 ± 9 m/s in the y direction.  Total irradiation times for the cumulative dose delivery for the 15 Gy, 99nA case was 792 ms and agreed within 0.1\% with the simulation and 0.01\% with the log file. For the data accumulated during the dose rate linearity study the total irradiation times agreed within an average 0.2 ± 0.2 ms when compared with the log file recordings and within 1 ± 1 ms compared to the simulated runs.

\subsection{Dose and Dose Rate Agreement}
\begin{figure}[ht]
\begin{center}
\includegraphics[width=16.5cm]{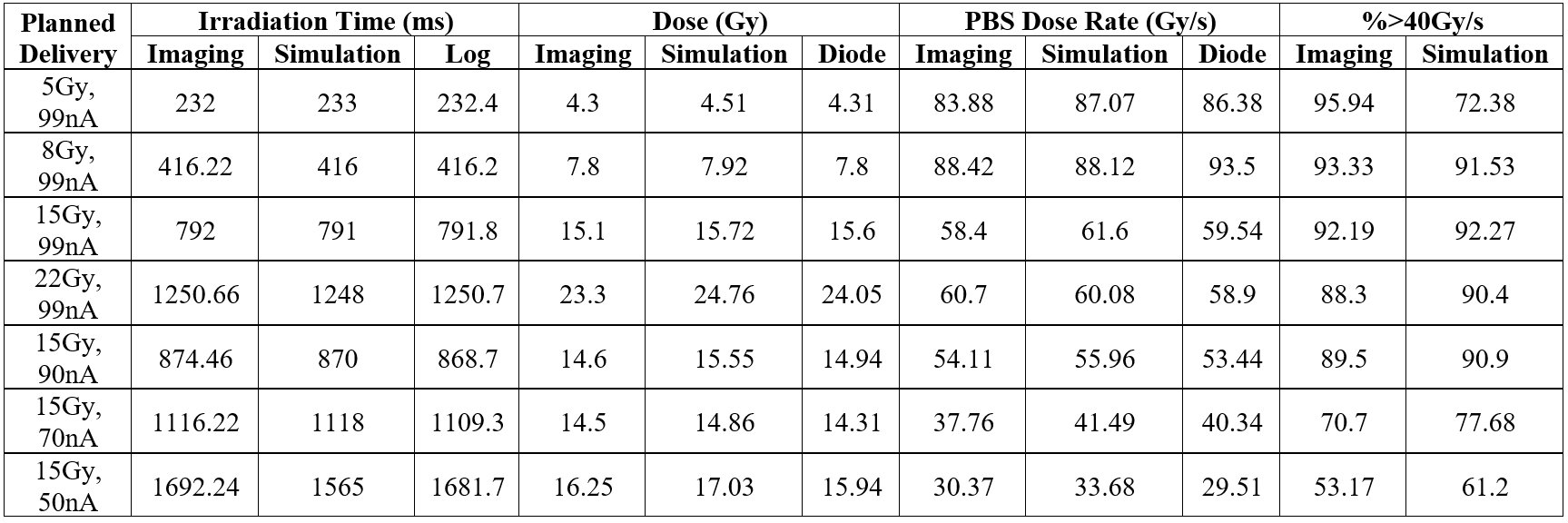}
\caption{Example of potential comparison metrics between the image data, simulation, log file recordings, and diode, highlighting the strong agreement across the data modalities.}
\label{fig_example3} 
    \end{center}
\end{figure}

\begin{figure}[ht]
\begin{center}
\includegraphics[width=16.5cm]{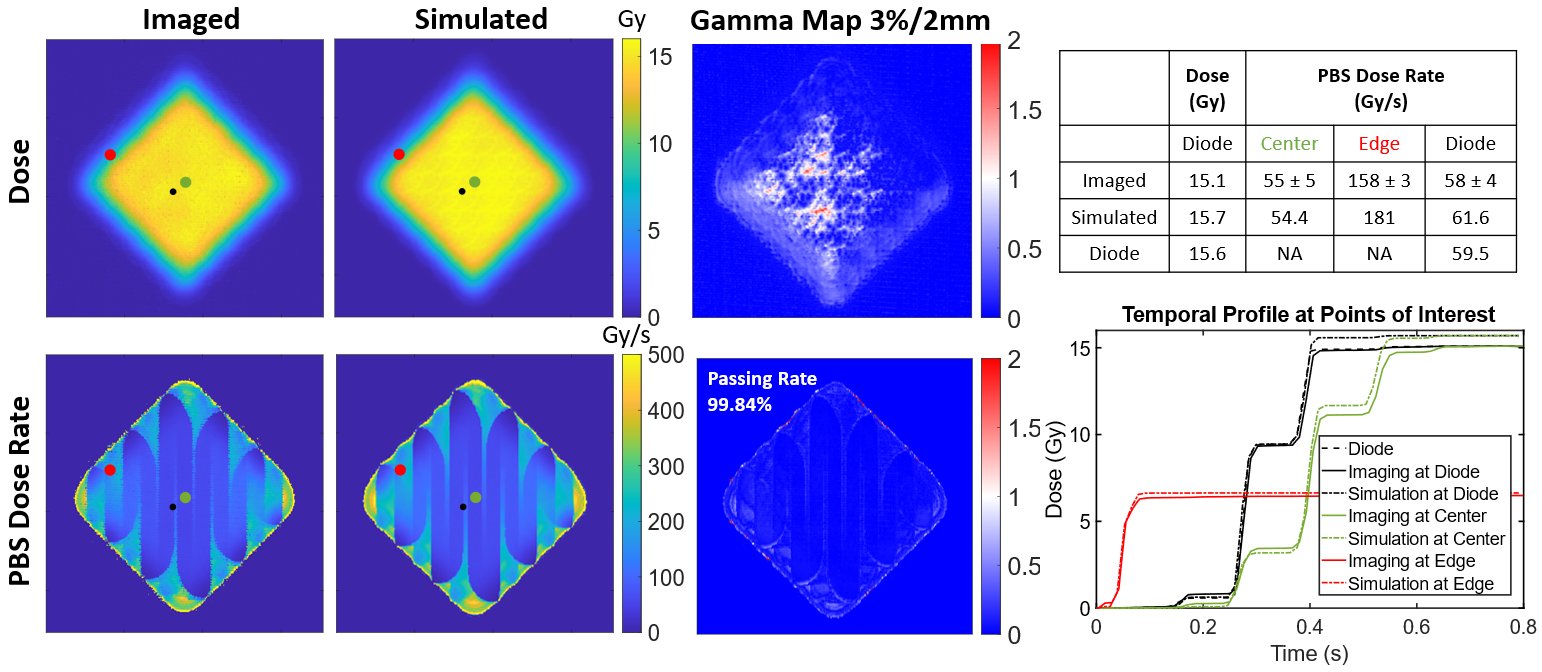}
\caption{Comparison of full-field dose and dose rate maps between imaging and simulated 15Gy delivery at 99nA, with temporal and representative dose rate measurements at three points of interest within the field.}
\label{fig_example4} 
\end{center}
\end{figure}

Calculating PBS dose-rate distribution is non-trivial due to complicated spatio-temporal dynamics of PBS delivery. Here we compare the dose rate quantities (see Table included in Figure \ref{fig_example3}) as reported by 1.) the scintillator camera system, 2.) the log-based dose rate value, 3.) the UHDR beam simulation, and 4.) the EDGE diode point detector in Table 1. For each field used in the dose and dose rate linearity studies, full-field dose and dose rate maps were compared between the imaging data and simulation, as shown with the example 15 Gy, 99nA case in Figure \ref{fig_example4}. To compare to the diode, an image was taken of the diode location without the scintillation sheet placed on top and under standard lighting conditions, allowing localization of the sensitive area of the diode within the images. \\

\indent The dose and dose rate values at the diode location, the center of the field, and the edge of the field (where higher dose rates are expected due to shorter delivery times), were compared to highlight the sensitivity of dose rate to the location of the field due to the scanning beam. Between the diode location and center of the field, though less than 1 cm apart, there are significant differences in the spatial temporal dose distribution, resulting in different calculations of PBS dose rate (55 vs 58 Gy/s or 5\% deviation). This highlights the importance of full-field information for better understanding of the dose rate distribution, as is elaborated on in the discussion. 
Cumulative 2D dose maps agreed with film measurements and simulation cumulative maps with a mean gamma passing rate at 3\%/2mm of 100\% and 99.3\%, respectively, thus confirming the spatial accuracy of the imaging system. The PBS dose rate maps agreed with the simulation with a mean gamma passing rate at 3\%/2mm of 98.7\%. The PBS dose rate measured at the diode location agreed between the imaging and the diode within 3±2\% and with the simulation within 5±3\%. The simulation agreed with the diode within 5±4\%. Mean dose rates were calculated with imaging system to be 18.7±2 Gy/s from the measured cumulative dose and total irradiation time and agreed with the log file within 2±1\%, with the simulation within 6±3\%, and the simulation agreed with the log files within 5±4\%. The percentage of the field receiving greater than 40Gy/s measured with the imaged maps agreed with simulation 8±8\%. \\

\subsection{PBS Dose Rate Thresholds}
\begin{figure}[ht]
\begin{center}
\includegraphics[width=15cm]{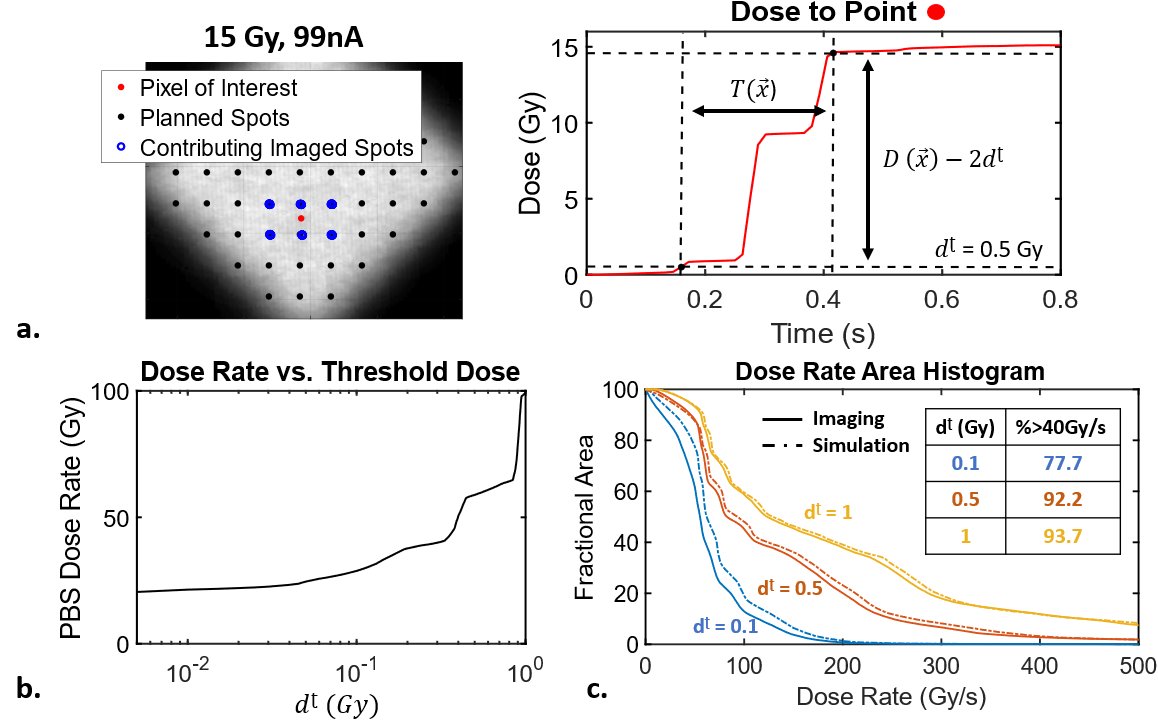}
\caption{a.) Visualization from imaging data of Folkert et al PBS dose rate definition with 0.5Gy $d_{\dagger}$. b.) Demonstration of dependency of PBS dose rate at the center of the field on the dose threshold, $d_{\dagger}$. c.) Dependency of the percentage of the field achieving the FLASH dose rate threshold on the dose threshold, $d_{\dagger}$.}
\label{fig_example5} 
\end{center}
\end{figure}

As mentioned in Section \ref{subdoserate}, dose rate, and thus the percent of the field receiving a certain dose rate, is highly dependent on input threshold parameters by the user. When defining PBS dose rate, Folkerts et al. simulated the dose rate dependency on $d_{\dagger}$, or the dose threshold at which the accumulation of dose is recognized, as shown in Figure \ref{fig_example5}. Here, we present the first validation of this simulation, where PBS dose rate is demonstrated to range from 35 Gy/s for a $d_{\dagger}$ of 0.01 Gy compared to 100 Gy/s for $d_{\dagger}$ equal to 1 Gy, as shown in Figure \ref{fig_example5}. Because of the clear dependency of dose rate on this threshold dose, it can be expected that the reported the percentage of the field achieving the FLASH dose rate threshold, which was set at 40 $Gy/s$ for this study to mimic the dose rate threshold set in FAST-01, would also be highly dependent on this dose threshold. Figure \ref{fig_example5} demonstrates that dependency, with deviations of 16\% between the percentage of fields, when only changing dose threshold. 

 \section{Discussion}
 Concurrent sub-millimeter ($<$0.5 mm) and sub-millisecond (±1 ms) spatiotemporal resolution is a unique capability of this camera-scintillator system, validated here. Additionally, this system enables remote, passive, real-time beam monitoring without interfering with the beam path. Existing, novel tools that have demonstrated the ability for temporal characterization of UHDR PBS systems include scintillating point detectors, diodes, and calibrated ionization chambers ; however, these systems lack spatial information important to the dynamics of PBS deliveries\cite{kanouta_time_2022,mcmanus_challenge_2020,rahman_characterization_2023}. The 2-D strip ionization chamber proposed by Yang et has reported similar spatial-temporal mapping of UHDR PBS deliveries at a sampling rate of 20 kHz, but requires implementation of 2D dose reconstructions and a fixed geometry setup \cite{yang_2d_2022}. Therefore, this system is the only available to provide external validation of UHDR PBS deliveries. \\
\indent The strongly inhomogeneous dose rate maps typical of UHDR PBS fields require high spatial resolution for quantitative assessment. With delivery times on the timescale of milliseconds to seconds, and spot dwell times on the order of milliseconds, maximum instantaneous dose rates were measured to be over 450 Gy/s in the case of 15Gy delivered at 99nA. This means, that when calculating PBS dose rate, a small deviation in temporal accuracy may lead to a high deviation in calculated PBS dose rate. The highest current, low dose case (5 Gy, 99nA) delivered with the shortest measured spot dwell times, at 3.8 ± 0.5 ms. Therefore, to accurately image this case with $<$3\% error in dose rate, assuming accurate capture of dose distributions, a dosimetry system would need to operate at over 10 kHz. This resolution requirement is highlighted by looking at the results of the 15Gy/99nA delivery again, where even a deviation by 0.001 seconds for the measured $T(\vec{x})$, as described in Figure \ref{fig_example4}, can lead to a deviation in the calculated PBS dose rate at the center of the field of 3.5 Gy/s or 7.5\%. Not only do UHDR-capable beams require dosimeters with high spatial resolution for full characterization, but they also require concurrent high temporal resolution, such as that provided by this imaging system.\\
 \indent Spatio-temporal agreement between the imaging system and various independent measurement devices was found to be within 5\%, confirming the accuracy and utility of ultra-fast imaging for future dose and dose rate evaluation of UHDR PBS beams. The ability to resolve per-spot and inter-spot information of the scanning beam at kHz repetition rate is extremely useful for estimating parameters, such as spot scanning speed and spot dwell time, which may inform treatment planning algorithms for further UHDR studies. The spot scanning speeds here reported agree within reasonable measurement resolution with that previously published by Kanouta et al who reported  10 ± 0.8 m/s and 25. 5 ± 5.1m/s estimates and that from Poulsen et al who measured 7 and 32 m/s \cite{kanouta_time_2022,poulsen_efficient_2018}, in the same respective directions.\\
\indent This work also demonstrates the importance of dose thresholds used in dose rate estimations, something currently not of convention in FLASH-RT research. While using Folkert et al.’s PBS dose rate definition, we noted a strong dependence of resulting dose rate quantities on the chosen threshold. To our knowledge, this is the first time the relationship between dose rate and the parameters of Equation \ref{eq:1} has been demonstrated with measured PBS field data. Unfortunately, the exact value of $d_{\dagger}$ is rarely used in FLASH dose rate reports in existing literature. The chosen $d_{\dagger}$ depends not only on thresholds applied during post-processing, but also will be limited by detector sensitivity and additional thresholds applied at lower level in dosimeter readout electronics. Going forward, such dose thresholds should be thoroughly assessed and reported, especially in the pre-clinical and clinical studies to establish consistency and comparability of the biological outcome-based results across different centers. Especially in the event some dose rate is needed to elicit the FLASH effect, treatment planning results will be highly dependent on this choice. In lieu of radiobiological data indicating the impact of the $d_{\dagger}$ threshold on the presence of the FLASH effect, we recommend investigators set $d_{\dagger}$ depending on the resolution of their measurement system. For this work, we found that a threshold of 50 cGy lead to repeatable measurements based on the dynamic range of our measurement system.
While the main limitation of this system is the restricted dynamic range, optimized here for UHDR deliveries, future work can be done to automate and extend this. Additionally, despite the small-buildup seen for shoot-through proton beams at such a shallow depth, some of the discrepancies between the PPC05/log/simulation and imaging and diode can be attributed to this buildup factor and the geometry of the required setup. Next steps to this work will involve additional comparison methods, improved system independence, and imaging of non-rigid surfaces. 
\section{Conclusion}
Spatio-temporal agreement with film and diode measurements confirms our ability to use this scintillation imaging system to accurately map dose rates of a UHDR PBS system. Additional comparison to the internally developed log-file-based reconstruction validated the potential to use this simulation for future study planning. Deviation of measured dose rate at the center, diode location, and edges of the field highlight the importance of full-field maps for UHDR PBS beam characterization, and dependency on dose thresholds in defining dose rate indicate the importance of clarifying and establishing standards for accurate comparisons across beams. Going forward, this camera system can be used to validate UHDR PBS beams via its unique, passive beam monitoring capabilities. This system is the only device available to provide concurrent high spatial and temporally resolved information of UHDR PBS deliveries.

\section*{Acknowledgments}
\indent The authors acknowledge SBIR R44CA268466, NIH R44 CA268466, and the Norris Cotton Cancer Center Shared Resources with NCI cancer center support grant 5P30 CA023108-41.

\section*{Conflict of Interest}
\indent This work has been funded by NIH grant R44 CA268466 and SBIR R44CA268466. Petr Bruza is affiliated with DoseOptics LLC, which provided hardware support for this study.





\printbibliography





\end{document}